\documentclass[prepint]{aastex62}

\usepackage[intlimits]{amsmath}
\usepackage{natbib}
\usepackage{xcolor}
\usepackage{subfigure}
\usepackage{hyperref}
\usepackage{url}
\usepackage{enumerate} 

\shorttitle{Clustering CMEs}
\shortauthors{Rodr\'iguez G\'omez et al.}

\begin{document}
\title{Clustering of fast Coronal Mass Ejections during the solar cycles 23 and 24 and implications for CME-CME interactions}

\correspondingauthor{Jenny Marcela Rodr\'iguez G\'omez}
\email{J.RodriquezGomez@skoltech.ru}

\author[0000-0002-9860-5096]{Jenny M. Rodr\'iguez G\'omez}
\affil{Skolkovo Institute of Science and Technology, Territory of innovation center “Skolkovo”, \\
Bolshoy Boulevard 30, bld. 1, Moscow 121205, Russia.}
\author{Tatiana Podladchikova}
\affil{Skolkovo Institute of Science and Technology, Territory of innovation center “Skolkovo”, \\
Bolshoy Boulevard 30, bld. 1, Moscow 121205, Russia.}

\author{Astrid Veronig}
\affil{Institute of Physics \& Kanzelh\"ohe Observatory,
University of Graz. 8010 Graz, Universit\"atsplatz 5}

\author{Alexander Ruzmaikin}
\affil{Jet Propulsion Laboratory, California Institute of Technology, Pasadena, California, USA}

\author{Joan Feynman}
\affil{Helioresearch, Glendale, CA 91214, USA}

\author{Anatoly Petrukovich}
\affil{Space Research Institute, Moscow, Russia}



\begin{abstract}

We study the clustering properties of fast Coronal Mass Ejections (CMEs) that occurred during solar cycles 23 and 24. We apply two methods: the Max spectrum method can detect the predominant clusters and the de-clustering threshold time method provides details on the typical clustering properties and time scales. Our analysis shows that during the different phases of solar cycles 23 and 24, CMEs with speed $\geq 1000\ km/s$ preferentially occur as isolated events and in clusters with on average two members. However, clusters with more members appear particularly during the maximum phases of the solar cycles. Over the total period and in the maximum phases of solar cycles 23 and 24, about 50\% are isolated events, 18\% (12\%) occur in clusters with 2 (3) members, and another 20\% in larger clusters $\geq 4$, whereas in solar minimum fast CMEs tend to occur more frequently as isolated events (62\%). During  different solar cycle phases, the typical de-clustering time scales of fast CMEs are  $\tau_c=28-32\ hrs$, irrespective of the very different occurrence frequencies of CMEs during solar minimum and maximum. These findings suggest that $\tau_c$ for extreme events may reflect the characteristic energy build-up time for large flare and CME-prolific active ARs. Associating statistically the clustering properties of fast CMEs with the Disturbance storm index \textit{Dst} at Earth suggests that fast CMEs occuring in clusters tend to produce larger geomagnetic storms than isolated fast CMEs. This may be related to CME-CME interaction producing a more complex and stronger interaction with the Earth magnetosphere.

\end{abstract}

\keywords{Sun: corona, coronal mass ejections (CMEs)}

\section{Introduction} 
\label{sec:intro}
Coronal mass ejections (CMEs) are a manifestation of solar variability and the main source of major space weather events. CMEs eject substantial amounts of mass and magnetic flux from the Sun to the Heliosphere and cause disturbances in the interplanetary medium \citep{2009EM&P..104..295G}. The CME initiation and impulsive acceleration occur on time scales of a few minutes to several hours with a kinetic energy that may exceed $10^{32}\ ergs$ \citep{2011ApJ...738..191B,2006SSRv..123...13H,2000ApJ...534..456V,1996AIPC..382..426S,1984JGR....89.2639H}. 
CME speeds range from $< 100 \ km/s$ to $> 2000 \ km/s$, occasionally reaching up to $3500 \ km/s$  \citep{2012LRSP....9....3W,2011LRSP....8....1C,2009EM&P..104..295G,2004JGRA..109.7105Y}. During their propagation, the interplanetary manifestations of the CMEs (ICME) may interact with the Earth (and other planets), producing space weather impacts on their environment and technology \citep{2018SSRv..214...21R,2013SpWea..11..585B,2006LRSP....3....2S}. 

The characteristics of extreme solar phenomena and extreme space weather events help us to better understand the dynamics and variability of the Sun as well as the physical mechanisms behind these events \citep{2017SSRv..212.1137K,2018SSRv..214...46G}. In this article, we will characterize extreme events in the form of fast CMEs. Large flares and fast CMEs predominantly originate from complex active regions that contain large amounts of magnetic flux \citep{2019LRSP...16....3T,2018SoPh..293...60M,2018ApJ...853...41T,2000ApJ...540..583S}. Observations have shown that active regions tend to occur in clusters. This behavior is related to the magnetic flux emergence of new active regions, which preferably emerge in the vicinity of old ones \citep{2011JGRA..116.4220R,Ruzmaikin1998,1983ApJ...265.1056G,1993SoPh..148...85H}. 

Multiple CMEs launched from complex active regions are not rare. \citet{2011JGRA..116.4220R} showed that fast CMEs in particular tend to occur in clusters. This clustering may lead to interactions of successive CMEs, either already close to the Sun or in interplanetary space. Solar observations reveal that CME-CME interaction occurs in particular for CMEs that are launched in sequence from the same active region. During their propagation from the Sun to Earth, the CME-CME interaction can be related to enhanced particle acceleration and can generate more intense geomagnetic storms than isolated CMEs when arriving at Earth \citep{2017SoPh..292...64L,2016SoPh..291.1447V,2015SoPh..290..579D,2004AnGeo..22.3679F}.

Here we study the temporal distribution of fast CMEs, with main focus on the statistical interpretation of the occurrence and clustering properties of CMEs, how these changes for different solar cycle phases, and also how the clustering properties may affect the CME's geo-effectivity. To this aim, we apply two different approaches, the Max Spectrum method and the de-clustering threshold time \citep{2006math......9163S, 2011JGRA..116.4220R, ferro2003}.
The Max Spectrum method provides two exponents: the power tail exponent ($\alpha$) describing the probability distribution of the speeds of fast CMEs and the extremal index ($\theta$) that separates individual clusters and also provides an estimate of the predominant cluster size. The de-clustering threshold time is used to identify clusters in time series of CMEs with speeds larger than $1000\ km/s$. This method provides information about the cluster size, the mean cluster duration and the mean time between successive fast CMEs.

The paper is structured as follows. In Section 2, we describe the CME data set. In section 3, we explain the  Max Spectrum method and the concept of de-clustering threshold time. In Section 4, the main results of our study on the clustering of fast CMEs that occurred during solar cycles 23 and 24 are presented. In addition, the same analysis is performed separately for different phases of the solar cycles. In Section 5 we present a statistical approach to relate the CME clustering properties to their potential geo-effectivity. In Section 6, we summarize our main findings and discuss their implications.

\section{Data set}\label{sec:data}

We use data from the Large Angle and Spectrometric Coronagraph\footnote{\url{https://cdaw.gsfc.nasa.gov/CME_list/}} (LASCO; \cite{1995SoPh..162..357B}) onboard the Solar and
Heliospheric Observatory (SOHO) satellite. This catalog contains all the CMEs detected by the LASCO coronagraphs C2 and C3, which cover a combined field of view from $2.1$ to $32 \ R_{\odot}$. The catalog provides several attributes to characterize the CMEs: date and time of first appearance in the coronagraph field of view, angular width, speed from the linear fit to the height-time measurements, speed from the quadratic fit at the last height of measurement, speed from quadratic fit at $20 \ R_{\odot}$ \citep{2009EM&P..104..295G,2004JGRA..109.7105Y}.
We use the speed from the quadratic fit at the time/distance of the first data point, which gives the speed closer to the initiation of the CME eruption in the low corona before other interactions occur.

We follow the approach outlined in \cite{2011JGRA..116.4220R} and build the time series from the hourly spaced time series of CME speeds. The hours with no CME occurrence are assigned a zero speed. In the few cases where more than one CME occurred within one hour, the highest CME speed is chosen. We use the entire LASCO data available from January $1996$ to March $2018$ (resulting in a total set of 25895 CMEs) covering almost completely the last two solar cycles. We note that our data also include the recently occurring strongest events of cycle 24, namely the X9.3 and X8.2 flare/CME events from 2017 September 6 and 10 (e.g. \cite{2017SoPh..292..131Y,2018ApJ...869...69M,2018SpWea..16.1156G,2018ApJ...852L...9S,2018ApJ...868..107V,2018ApJ...867L...5L,2019SoPh..294....4R}). 

\begin{figure}[!ht]
   \centering
   \includegraphics[width=\textwidth]{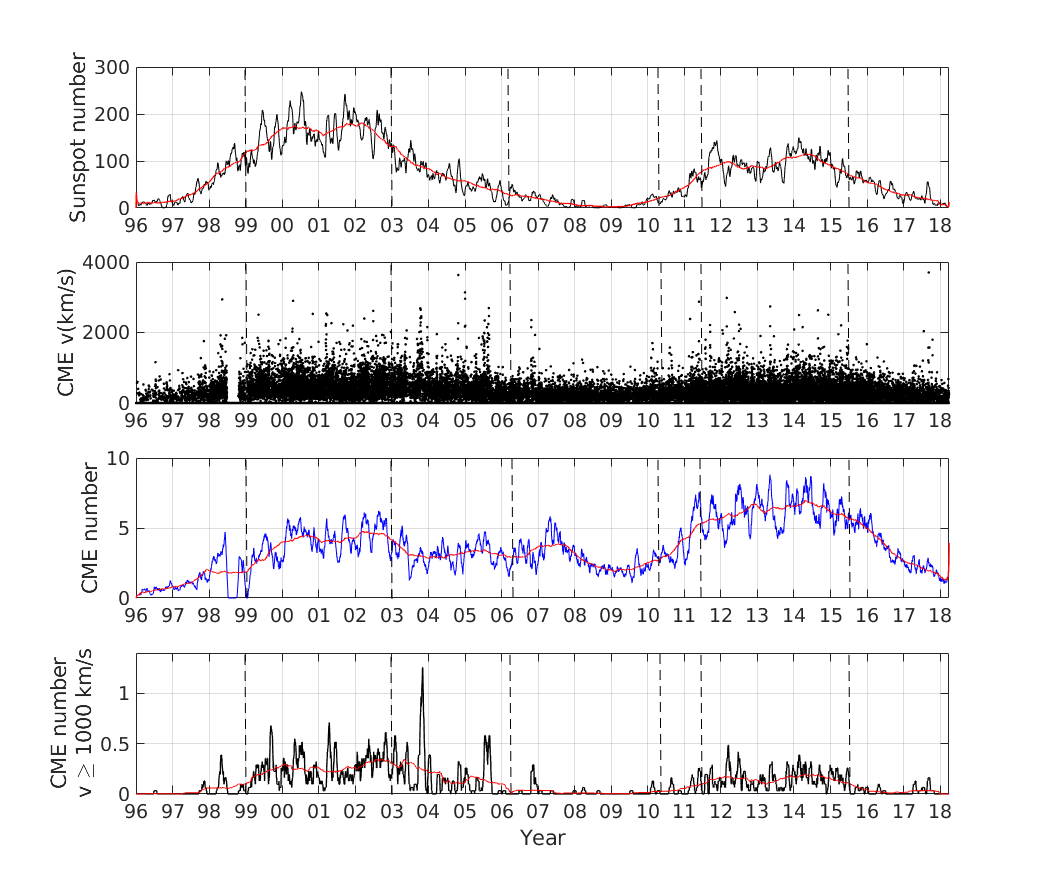}
  \caption{Top: Monthly mean Sunspot Number. Second: Time series of the CME speed. The data gap from June to November 1998 is due to the intermediate loss of contact with the SOHO satellite. Third: Monthly mean of the daily number of CMEs. Bottom: Monthly mean of the daily number of CMEs with speeds $v\geq 1000 \ km/s$. The data  cover the time range from 1996 to 2018, with the red lines showing the 13-month running mean.  
  The analysis is carried out for the full range shown. Vertical dotted lines mark the sub-intervals selected for separate analysis.}
  \label{fig:figcmes}
   \end{figure}

Figure \ref{fig:figcmes} shows the CME speeds and their number together with the sunspot number from 1996 to 2018. The monthly mean sunspot number was obtained from  WDC-SILSO, Royal Observatory of Belgium (top panel). The second and third panel show the CME speed and the monthly means of the daily number of CMEs from the CDAW LASCO Catalog.
The monthly mean of the daily number of CMEs with speeds $v\geq 1000 \ km/s$ are plotted in the bottom panel. These curves were smoothed over 13 months (red lines). The vertical dotted lines mark three 4-year intervals centered at the maximum and minimum phases of  solar cycles 23 and 24, for which we study the CME clustering properties also separately. 

\begin{figure}[!ht]
   \centering
   \includegraphics   [width=0.7\textwidth,height=0.4\textwidth]
   {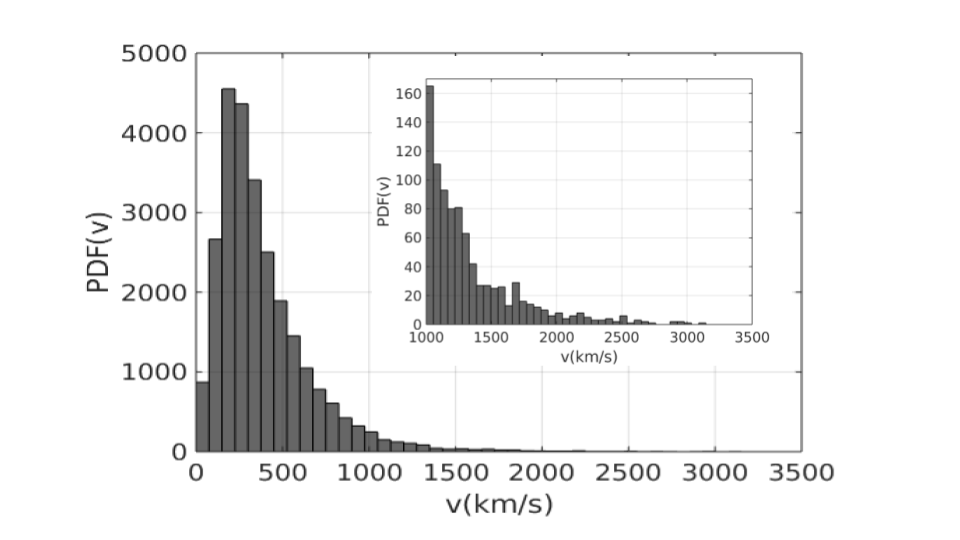}
  \caption{Distribution of the CME speed from SOHO/LASCO catalog for solar cycles 23 and 24. The inset shows a close-up of the CME speed distribution for $v\geq 1000\ km/s$.}
  \label{fig:cmeswhole}
   \end{figure}

Figure \ref{fig:cmeswhole} shows the Probability Distribution Function (PDF) for the solar cycles 23 and 24 and a close-up of the distribution for speeds $v\geq1000\ km/s$. These PDFs of the CME speeds reveal a non-Gaussian heavy-tailed distribution \citep{2011JGRA..116.4220R, 2005ApJ...619..599Y}. This means that fast CMEs occur with a much higher probability than expected from a normal or exponential distribution. The distribution of CMEs speeds shown on top has the mode at $195\ km/s$, the mean speed is $357 \pm 282 \ km/s$. $899$ CMEs ($3.5\%$ of the overall sample) have speeds exceeding $1000 \ km/s$. $2847$ CMEs ($11\%$) have speeds $> 700\ km/s$, and $63$ CMEs ($0.2\%$) achieves speeds $> 2000\ km/s$.

The dominant interplanetary phenomena causing intense magnetic storms ($Dst < -100 \ nT$) are the interplanetary manifestations of fast coronal mass ejections (ICMEs). The statistical dependence of \textit{Dst} minima during storms were widely explored \citep{2018JSWSC...8A..22P,2012SpWea..10.7001P,2013JGRA..118..385E,2008JGRA..113.5221E,2007JASTP..69.1009K}.
The main properties that determine the geo-effectivity of an ICME impacting the Earth magnetosphere are its arrival speed and the strength of the $B_{z}$ component of the interplanetary magnetic field \citep{2006JGRA..111.7S08B, 1999SSRv...88..529G, 1991JGR....96.7831G}. While currently we still have no proper handle to derive estimates of $B_{z}$ from solar observations (e.g. \citet{2019RSPTA.37780096V,2018SSRv..214...46G}), the ICME speed impacting at Earth (though evolving in interplanetary space due to the drag force exerted by the ambient solar wind) is related to the CME speed that we can derive from the coronagraph  observations near the Sun (e.g. \citet{2013SoPh..285..295V}). Thus, in order to characterize extreme events, we focus our analysis here in particular on fast CMEs, defined with speeds $v>1000 \ km/s$, which comprise about 3.5\% of the overall sample.


\section{Methods} \label{sec:method}
Here we describe two statistical methods to study and characterize the clustering properties of the fast CMEs. These are the Max Spectrum and the de-clustering threshold time method.

\subsection{Max Spectrum method}
The Max Spectrum method is based on the block maxima technique \citep{2006math......9163S}. In general a real valued random variable $X$ with a cumulative distribution function $F(x)=P\{X\leq x\}$, $x\in R$ is said to have a right heavy tail if, 
\begin{equation*}
 P\{X>x\}=1-F(x)=L(x)x^{-\alpha}, \ as \ x \rightarrow \infty
\end{equation*}
for some $\alpha>0$, where $L(x)>0$ is a slowly varying function. The tail exponent $\alpha>0$ controls the rate of decay of F and hence characterizes its tail behavior. If we consider the case where the slowly varying function $L$ is trivial, when

\begin{equation}
 P\{X>x\}=1-F(x)\sim \sigma_{0}^{\alpha}x^{-\alpha}, \ as \ x \rightarrow \infty
 \label{eqj}
\end{equation}
with $\sigma_{0}$ and where $\sim$ means that the ratio of the left-hand side to the right-hand side in Eq.\ref{eqj} tends to $1$, as $x \rightarrow \infty$. We assume that the $X(i)'s$ are almost surely positive ($F(0)=0$) \citep{2006math......9163S}.

In the application of this method, we use the hourly times series of CME speeds  created, without using a CME speed threshold. In our case the variable $X(i)$ corresponds to the CMEs speed $v(i)$.
This method starts with taking averages of data maxima in time intervals (blocks) with a fixed size. The block size is then progressively increased.
Here we consider the time series of total length N for the CME speed $v(i)$, where $1\leq i \leq N$ and $j$ introduce the time interval scale index $j=1,2,3,\dots log_{2}(N)$. To form non-overlapping time blocks of length $2^{j}$, 
at each fixed scale $j$ we calculate the maximum CME speed within each block
\begin{equation}
 D(j,k)=\max_{1\leq i \leq 2^{j}} v(2^{j}(k-1)+i) \hspace{1 cm}   k=1,2,\dots b_{j+1}
\end{equation}

where $b_{j}=\left[\frac{N}{2^{j}}\right]$ is the number of blocks (of length $2^{j}$) at each scale $j$ and $k$ defines the location of the block on the time axis. 


The blocks of scale $j$ are naturally contained in the blocks of scale $(j+1)$. Now, we average the binary logarithms of the block maxima $D(j,k)$ over all blocks at a fixed scale $j$, i.e.
\begin{equation}
 Y(j)=\frac{1}{b_{j}} \sum^{b_{j}}_{k=1} log_{2} D(j,k)
\end{equation}
 The function $Y(j)$ is called the Max Spectrum of the data. \citet{2006math......9163S} established an important result: The Max Spectrum for time series with  sufficiently large $j$ scales can be expressed as
\begin{equation}
 Y(j)\simeq \frac{j}{\alpha} + C
 \label{eq}
\end{equation}

where $C$ is a constant and $\alpha>0$. The tail of the data distribution follows a power law with exponent -$\alpha$. The exponent $\alpha$ is called the power tail exponent, which allows us to characterize what kind of distribution is related to the time series. In general, under the Generalized Extreme Value (GEV) 
theory, some distributions are obtained depending on location and a scale parameter. One of them is called the Fr\'echet distribution. In general, this distribution shows a right-side tail that decays like a power law \citep{macneil2005,GVK491731736,Hsing1988,1983Leadbetter}. \cite{2011JGRA..116.4220R} have shown that fast solar CMEs can be described as a Fr\'echet distribution. In this paper, we are interested in the Fr\'echet distribution and parameters describing the clustering of the fast CMEs that occurred during solar cycles 23 and 24. 
 

Eq. \ref{eq} is valid for statistically independent events. If we have dependent data with the same distribution function \citep{2006math......9163S}, then, Eq. \ref{eq} can be transformed into 
\begin{equation}
 Y(j)\simeq \frac{j}{\alpha} + C +log_{2}\left(\theta\right)/ \alpha
 \label{eq1}
\end{equation}

where $C$ is a constant, $\theta$ is called \textquotedblleft the extremal index\textquotedblright which takes values in the interval $[0,1]$ \citep{1983Leadbetter}. Values of $\theta$ close to $0$ indicate a strong dependence and the possibility to form clusters, while values close to $1$ show weak dependence indicating individual independent events. Note, this index characterizes only the dependence of the {\em extremes} in the time series data \citep{2010arXiv1005.4358H}.

Eqs. \ref{eq} and \ref{eq1} suggest a method of estimating $\alpha$ and $\theta$ \citep{2010arXiv1005.4358H,2006math......9163S}. 
The power tail exponent $\alpha$ is obtained on the self-similar range of the Max Spectrum $Y(j)$. This range can be related to the self-similar cascade process in turbulence \citep{2011JGRA..116.4220R,1995tlan.book.....F}. Similarly, here we check the self-similar interval to obtain the slope of the line fitted and obtain the power tail exponent $\alpha$. The inverse exponent $1/\alpha$ is obtained as a slope of the line fitted to the Max Spectrum of the data in the self-similar range \citep{2011JGRA..116.4220R,2006math......9163S}. The process to select the self-similar range is fundamental to obtain the power tail exponent and it influences the extremal index $\theta$ and the cluster number. In general, different intervals were checked on the self-similar range to obtain the slope and  the power tail exponent ($\alpha$). The best line fitted and their corresponding correlation coefficient guides the choice of this interval. 
The Max Spectrum $Y(j)$ and the power tail exponent $\alpha$ are key parameters in the estimation of the extremal index $\theta$.


The extremal index ($\theta$) defines the number of independent clusters and provides an estimate of the cluster size given as ($1/\theta$) \citep{1983Leadbetter}.
To calculate the extremal index $\theta$, the original data is first randomly permuted. The new data series ($v_{i}^{*}$) is obtained in the interval $1\leq i \leq N$, and has the same distribution function as the original data. The randomization destroys the dependence structure of the data, resulting in an approximately independent sample \citep{2010arXiv1005.4358H}.
The new Max Spectrum $Y(j)^{*}$ is related to the data series randomly permuted, and satisfies Eq. \ref{eq}, which means the tail of the new data distribution follows a power law with exponent -$\alpha$. Since the Max Spectrum of the original data $Y(j)$ satisfies Eq. \ref{eq1} with the same constant $C$, the difference between the two spectra yields an estimate of the extremal index
\begin{equation}
 \theta=2^{-{\alpha}(Y(j)^{*}-Y(j))}
 \label{the}
\end{equation}
where $\alpha$ is the fitting parameter in the power tail exponent obtained from the Max Spectrum $Y(j)$.
 Then, we calculate the differences of $Y(j)^{*}-Y(j)$ and compute the mean for positive differences to obtain an estimate of the extremal index $\theta$ at each scale $j$. This procedure is repeated $100$ times and 100 $\theta(j)$ values are calculated to produce a sequence of $\theta(j)$ boxplots for each scale $j$. We obtain an empirical $95\%$ confidence interval, based on $0.025th$ and $0.975th$ empirical quantiles of $\theta(j)$ from the histogram of $\theta$ values. 
 In practice, \citet{2010arXiv1005.4358H} recommend selecting the middle range of scales for $\theta$ estimation. At large scales $j$ (larger block sizes) the bias is lower and the number of block-maxima is reduced. At lower scales $j$ (smaller block sizes) the bias grows. 
 
We developed a code to calculate the Max Spectrum and extremal index $\theta$.
The performance of these estimations is examined carefully. We compared our results with the code\footnote{https://sites.lsa.umich.edu/sstoev/software/} of \citet{2006math......9163S} and the results presented in \citet{2011JGRA..116.4220R} for the years 1999 to 2006. Additionally, we examined our code using the Max-AutoRegressive model of order one (Max-AR(1)) to obtain the extremal index $\theta$. We use the Max-AR(1) series as $[y] = max\_ar ( b, a, iter ,N)$, with the length of time series $N=2^{15}$, number of iteration $iter=1$, and the parameters $a=0.7$, $b=1.5$. Our results are in agreement with the results based on the Stoev's code.\\ 

\subsection{De-clustering threshold time $\tau_{c}$}
In addition, to derive a more detailed characterization of the cluster properties and how they change over different phases of a solar cycle, we use the de-clustering threshold time method. The main idea is to identify clusters in the time series of CME with speed $v\geq 1000 \ km/s$. The mean time interval between CMEs within a cluster depends on the speed threshold ($v$). In our case, we choose CME speeds $v \geq 1000 \ km/s$ to characterize extreme events. 
The extremal index provides an estimate of the number of clusters ($m\times \theta$), where $m$ is the number of extreme events within a given time interval, e.g. $m$ CMEs with speeds exceeding a threshold $v$ occur during this interval. These CMEs are on average grouped into a cluster. The de-clustering threshold time $\tau_{c}$ concept is useful to group CMEs into clusters. Consider the time intervals $\tau_{i}$ between consecutive fast CMEs. If the time interval between two fast CMEs is less than $\tau_{c}$, then these CMEs can be grouped into a cluster \citep{2011JGRA..116.4220R,beirlant2004statistics,ferro2003,smith1989}.

\begin{figure}[!ht]
\centering
\subfigure[Solar cycles 23 and 24]{\includegraphics[width=0.35\textwidth, height=0.35\textwidth,keepaspectratio]{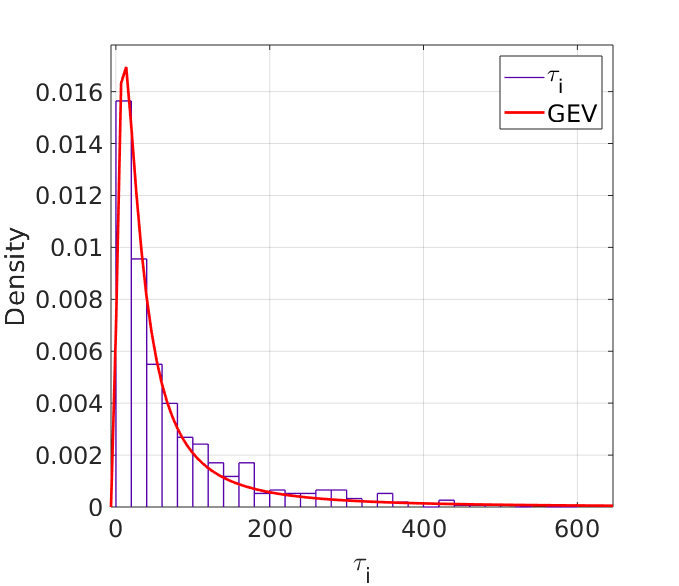}}
\subfigure[Maximum Solar cycle 23]{\includegraphics[width=0.35\textwidth,height=0.35\textwidth,keepaspectratio]{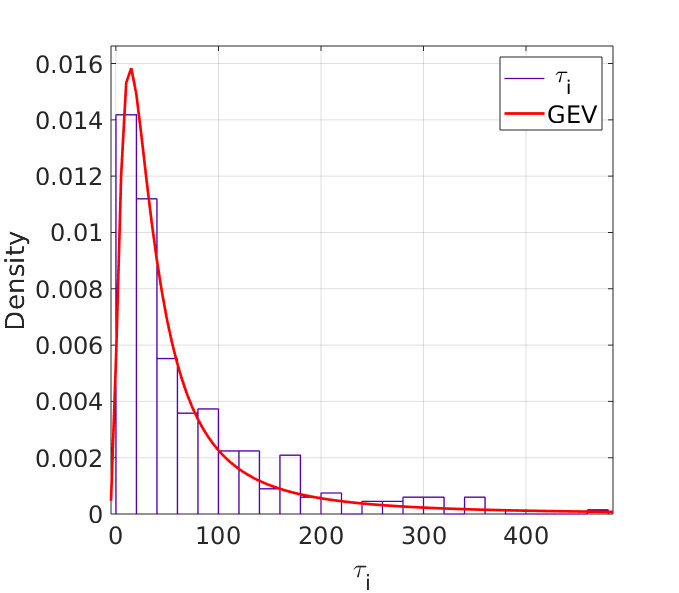}}
\subfigure[Maximum Solar cycle 24]{\includegraphics[width=0.35\textwidth,height=0.35\textwidth,keepaspectratio]{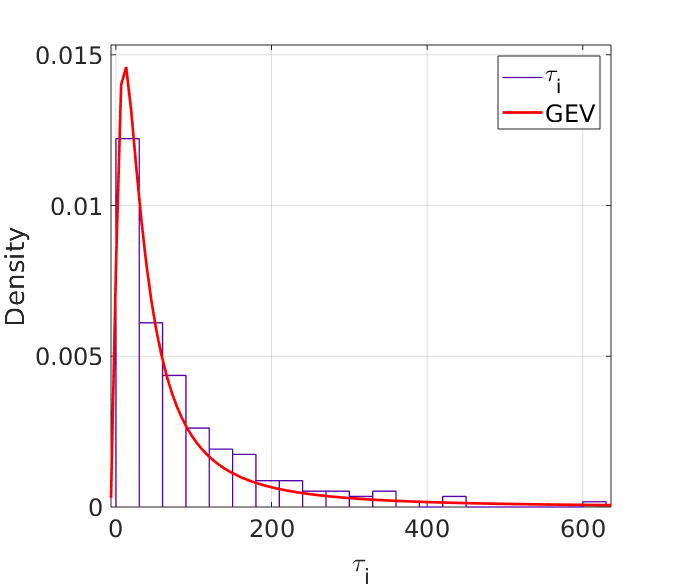}}
\subfigure[Solar minimum]{\includegraphics[width=0.35\textwidth,height=0.35\textwidth,keepaspectratio]{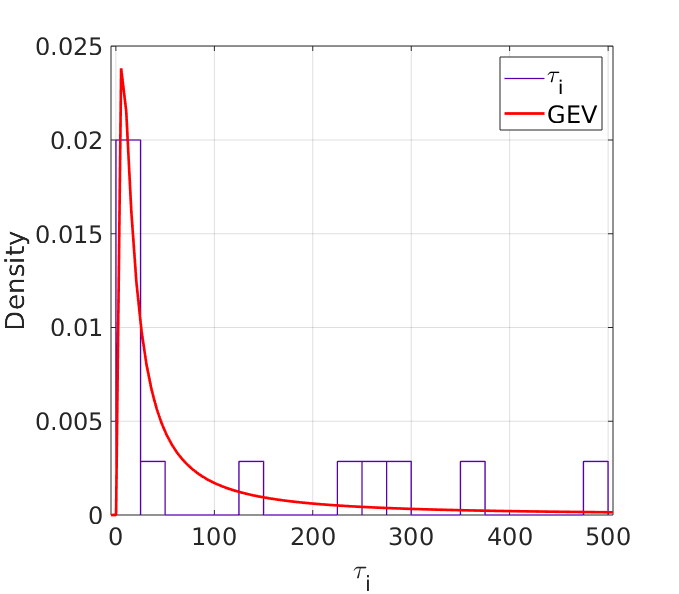}}
\caption{Determination of the de-clustering threshold time $\tau_{c}$ at each phase of the solar cycle. Probability Density Function (PDF) and Generalized Extreme Value (GEV) distribution (red line) of time intervals $\tau_{i}$ between consecutive fast CMEs. (a) Solar cycles 23 and 24, derived declustering time $\tau_{c}=28.0\ hrs$, (b) Maximum of solar cycle 23, $\tau_{c}=28.2 \ hrs$, (c) Maximum of solar cycle 24 $\tau_{c}=32.0\ hrs$, (d) Solar minimum $\tau_{c}=32.5 \ hrs$.}
\label{fig:tau}
\end{figure}

To determine the de-clustering threshold time $\tau_{c}$, we use the Probability Density Function (PDF) of time intervals $\tau_{i}$ between consecutive fast CMEs and the Generalized Extreme Value (GEV) distribution. $\tau_{c}$ was defined as the maximum ($\sigma$) value of the Generalized Extreme Value (GEV) distribution of time intervals ($\tau_{i}$) between consecutive fast CMEs.
Figure \ref{fig:tau} shows the Probability Density Function (PDF) of time intervals $\tau_{i}$ between consecutive fast CMEs ($v\geq 1000\ km/s$) and the Generalized Extreme Value (GEV) distribution (red line) at each period of the solar cycle. The de-clustering threshold time during the whole interval (solar cycles 23 and 24) is $\tau_{c}\leq 28.0 \pm 1.4 \ hrs$ (Figure \ref{fig:tau}, panel (a)). Applying the method separately to the different phase of the solar cycle, we find $\tau_{c}\leq 28.2 \pm 2.0 \ hrs$ for the maximum of cycle 23,  $\tau_{c}\leq 32.0 \pm 3.6 \ hrs$ for the maximum of cycle 24, and $\tau_{c}\leq 32.5 \pm 17.5 \ hrs$ for the minimum phase between cycles 23 and 24 (Figure \ref{fig:tau}, panels b-d).

\section{Clustering of the observed fast Coronal Mass Ejections}

Here we apply the methodology described in Section \ref{sec:method} 
to fast CMEs listed in the LASCO CDAW catalog. The Max Spectrum $Y(j)$ within the self-similar range is used to obtain the power tail exponent ($\alpha$). In the Max Spectrum plots, the $log_{2}$ units along the y-axis are converted to $km/s$, using $2^{Y(j)}$ and the scales $j$ on the x-axis are converted into time units using $2^{j}$. See the top panel in figures \ref{fig:whole} - \ref{fig:min}, which show the results for different periods during solar cycles 23 and 24.
We obtained a set of boxplots with $\theta$ values (middle panel figures \ref{fig:whole} - \ref{fig:min}). The central mark in  boxplots is the median, the box edges are the $10th$ and $90th$ percentiles and whiskers extend to the most extreme data points. Additionally, we built a histogram with the $\theta$ values and we calculate an empirical 95\% confidence interval, based on $0.025th$ and $0.975th$ empirical quantiles in each histogram (vertical dotted lines in the bottom panel Figures \ref{fig:whole} - \ref{fig:min}). This procedure allows us to obtain an interval of the extremal index $\theta$ values as well as to obtain an estimate of the predominant cluster size ($1/\theta$).

In Sect. 4.1 we present the analysis applied to the whole time period under study, i.e. from January 1996 to March 2018. However, the CME occurrence rate and speeds vary over the solar cycle (see Figure \ref{fig:figcmes}). To take this variability into account, in Sects. 4.2 - 4.4, we apply the same method for different sub-periods to study the variations of the CME clustering over different solar cycle phases. In particular, we select three periods each covering 4-years, representative of the maximum phase of cycles 23 and 24 as well as the minimum between cycles 23 and 24, as marked in the Figure \ref{fig:figcmes}.
\newpage
\subsection{Full period 1996 - 2018}
The full interval covers the time range from January 1996 to March 2018, i.e. it covers almost entirely cycles 23 and 24. During this period the length of the hourly CME speed time series is $N=193944$ and  the number of scales is $j=log_{2}(N)$. Thus, we have $j=17$ available scales to apply the Max Spectrum method. 
\begin{figure}[!ht]
\centering
\includegraphics[width=0.65\textwidth,height=0.65\textwidth,keepaspectratio]{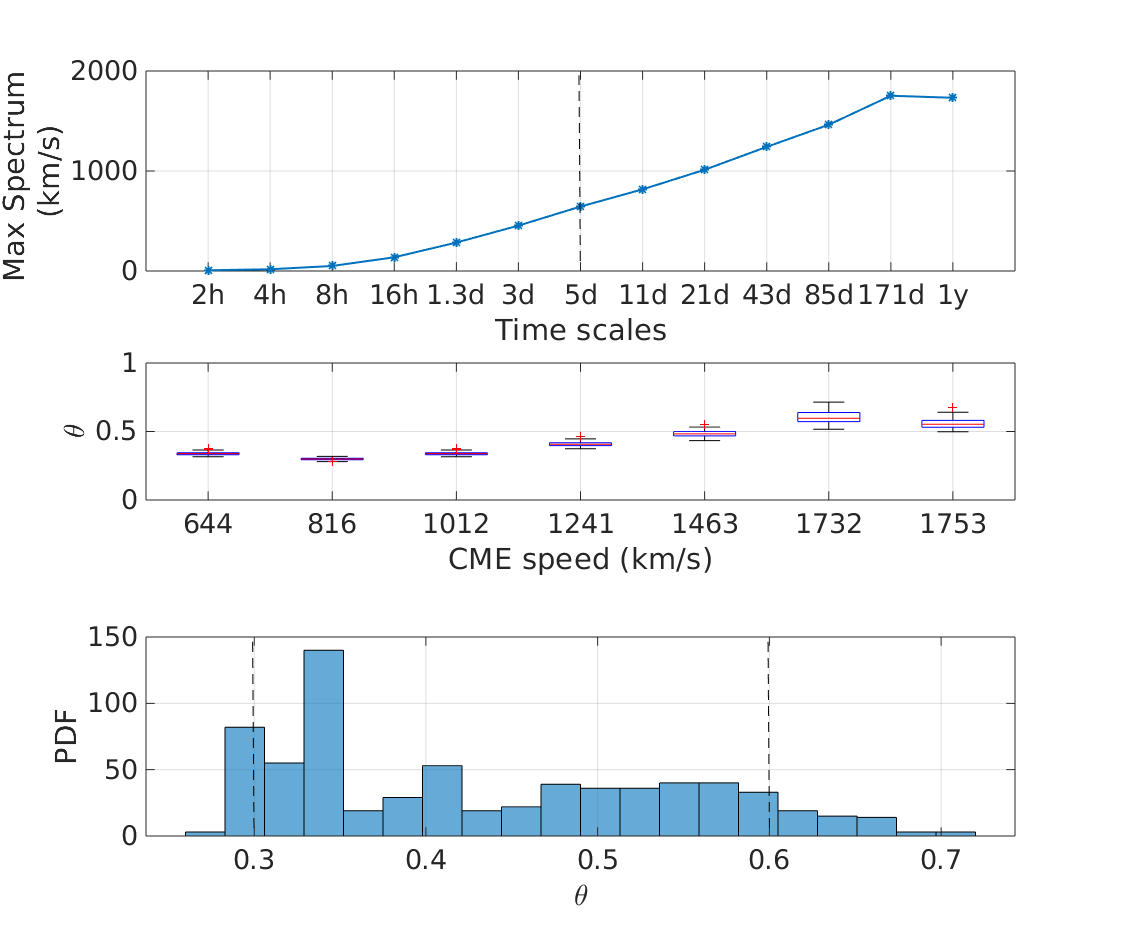}
\caption{The Max Spectrum method applied to the CMEs speeds during the full period covering the solar cycles 23 and 24.
Max Spectrum of the CME speeds at progressively increasing time scales. The vertical dotted line indicates the starting scale for the self-similar range (top). Boxplots of the extremal index ($\theta$) obtained by the Max Spectrum method the speed range of $644 \ km/s$ to $1753 \ km/s$ interval (middle). The histogram of the extremal index ($\theta$), the dotted vertical lines show the 95\% empirical confidence intervals (bottom).}
\label{fig:whole}
\end{figure}
Our best fit to the slope gives evidence that the cumulative distribution function of the CMEs speeds has a Fr\'echet type power law tail, with a power law exponent $\alpha =3.5$ (Figure \ref{fig:whole} (top)). Boxplots of the extremal index ($\theta$) in the scales related to the self-similar range are from $644 \ km/s$ to $1753 \ km/s$ (Figure \ref{fig:whole} (middle)).
The $0.025th$ and $0.975th$ empirical quantiles of the histogram of the $\theta$ boxplots (Figure \ref{fig:whole}, bottom) allow us to obtain an estimate of the extremal index $\theta$, which shows values from 0.36 to 0.66 within the 95\% confidence interval, with a mean of $\theta=0.43$. The corresponding cluster size is $1/\theta \approx$ 2 to 3, which means that in the whole time period CMEs with speeds higher than $644 \ km/s$ occur preferentially in groups of two or three. \\

When selecting CMEs with speeds $v\geq1000\ km/s$ we obtain $m=913$ extreme events, with $\theta\approx 0.36$ to $0.66$ and the estimated number of clusters is $\theta \times m \approx 328$ to $603$, with a de-clustering threshold time $\tau_{c}=28.0 \ hrs$ as derived from the maximum of the GEV fit (Figure \ref{fig:tau}). Table \ref{font} summarizes the derived information on the cluster size, mean cluster duration, mean time between successive CMEs and an estimate of the probabilities that a cluster of the corresponding size is recorded, using the de-clustering threshold time ($\tau_{c}$) method. The cluster duration was calculated as the time difference $\Delta t$ between the end and the start of the cluster, with the start time of the cluster being defined as the first appearance of the first CME of the cluster in the LASCO C2 FOV, and the end time defined as the first appearance of the last CME in the cluster in the LASCO C2 FOV.
In general, we find that about half of the events (49.2\%) occur as individual events, but CMEs in clusters with two and three members are also prominent, with a percentage of 18.4\% and 12.5\% respectively. However, also CMEs that occur in clusters with 4, 5, 6 and 7 members exist on a significant percentage, between about 3 and 7\%. 
The probability of recording a cluster consisting of one (isolated) fast CME is 0.742. The probability of recording a cluster with 2 or 3 fast CMEs within the de-clustering threshold time $\tau_{c} \leq 28\ hrs$ is 0.140 and 0.063. These probabilities describe the occurrence of a fast CME followed by one or two more fast CMEs. Finally, clusters with $\ge$4 members show a probability smaller than 0.055, see Table \ref{font}. 


\begin{table}[ht!]
\begin{center} 
\caption{CME clusters in solar cycles 23 and 24, with speeds exceeding $1000 \ km/s$ and  $\tau_{c}\leq 28.0 \pm 1.4\ hrs$. The first column gives the size of clusters. The second column lists the number of clusters of a certain size. The third column gives the total number of CMEs in these clusters and their percentages. The fourth column lists the mean duration of the cluster with the standard error in parenthesis. The fifth column lists the mean time between successive CMEs in each cluster with the standard deviation in parenthesis. The sixth column provides an estimate of the probabilities that a cluster of the corresponding size is recorded, which is the number of clusters of specific size divided by the total number of clusters.} 
\begin{tabular}{c c c  c cc}
\hline 
Cluster size& Number of clusters & Number of CMEs &Mean cluster & Mean time between& 
Recording\\
&  &in cluster (\%)&duration(hrs)&successive CMEs (hrs)&probabilities \\
\hline
1& 449&449(49.2)&-&-&0.742\\
2& 84&168(18.4)&13.5(1.0)&13.4(9.3)&0.140\\
3&38&114(12.5)&28.4(2.0)&13.7(8.3)&0.063\\
4&16&64(7.0)&39.2(5.0)&12.7(8.0)&0.026\\
5&5&25(2.7)&60.4(10.1)&13.1(7.2)&0.008\\
6&5&30(3.3)&59.4(11.4)&10.2(7.7)&0.008\\
7&5&35(3.8)&68.2(14.0)&12.7(8.2)&0.008\\
8&1&8(0.9)&81.0&10.5(7.2)&0.002\\
10&2&20(2.2)&133.7(27.1)&13.5(7.4)&0.003\\
\hline
Total&605&913(100)&-&-&1\\
\hline
\end{tabular}
\label{font}
\end{center}
\end{table}
In the following, we apply the same type of analysis separately to three sub-intervals, each of lengths 4 years, that characterize the maximum of cycles 23 and 24 and the minimum between then two cycles.

\subsection{Maximum of solar cycle 23}
\begin{figure}[!ht]
\centering
\includegraphics[width=0.65\textwidth,height=0.65\textwidth,keepaspectratio]{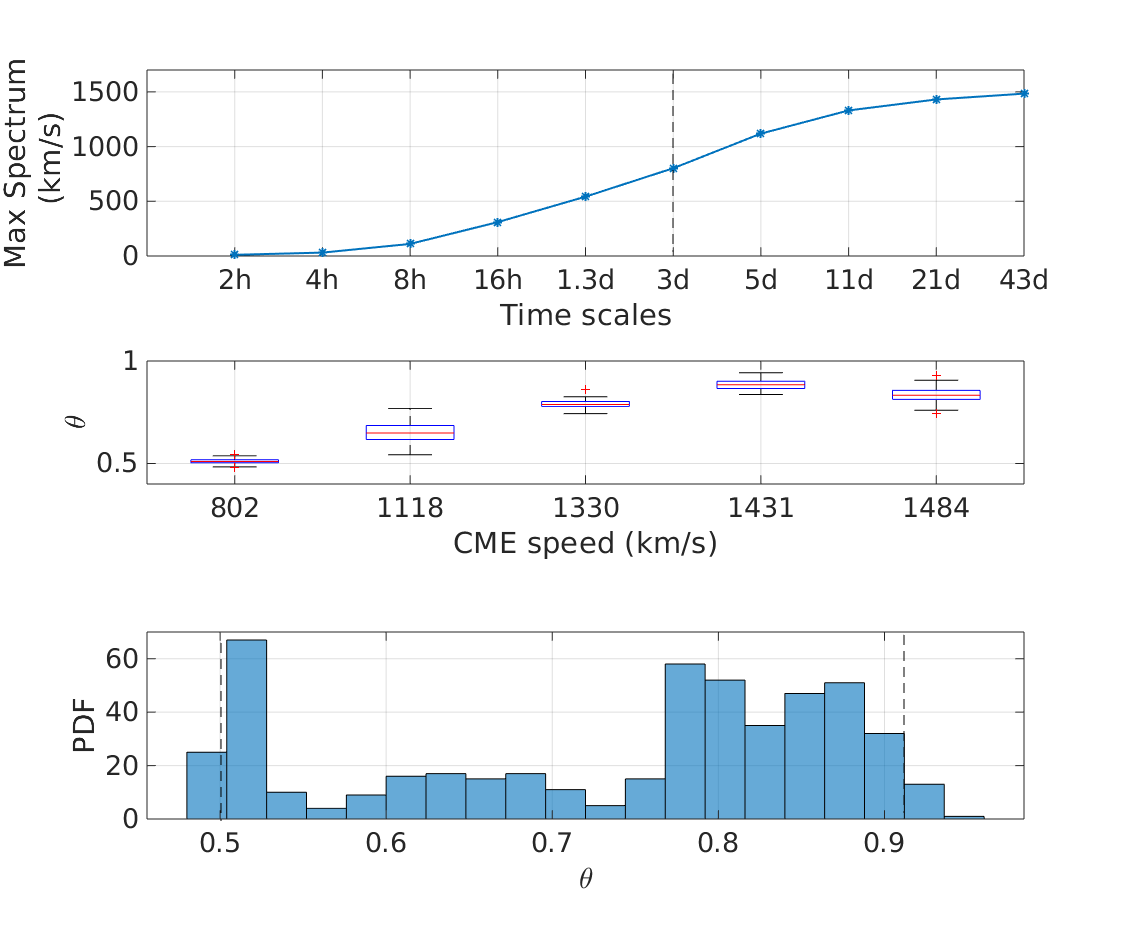}
\caption{The Max Spectrum method applied to the CMEs speeds during the maximum of solar cycle 23 at progressively increasing time scales. The vertical dotted line indicates the starting scale for the self-similar range (top). Boxplots of the extremal index ($\theta$) obtained by the Max Spectrum method the speed range of $802 \ km/s$ and $1484 \ km/s$
interval (middle). The histogram of the extremal index ($\theta$), the dotted vertical lines show the 95\% empirical confidence intervals (bottom).}
\label{fig:max23}
\end{figure}

To study the cluster behavior during the solar maximum of solar cycle 23, we select the data for the time range from February 1999 to February 2003. The length of this time series is $35736$ and $j=15$ scales are available in this interval. We find the Max Spectrum is self-similar in a range from $802 \ km/s$ to $1484 \ km/s$ and the power tail exponent $\alpha=2.8$ (Figure \ref{fig:max23} (top)). The extremal index $\theta$ ranges from 0.49 to 0.91, within the empirical 95\% confidence interval (middle and bottom panels figure \ref{fig:max23}) with a mean value of 0.73, corresponding to a cluster size of $1/\theta\approx$ 1 to 2. In this period, we have $m=377$ extreme events with $v\geq1000 \ km/s$. Using the extremal index values $\theta \approx 0.49$ to $0.91$, we obtain an estimated for the number of clusters $\theta \times m\approx 185$ to $343$.
For the de-clustering threshold time, we obtain $\tau_{c}\leq 28.2 \pm 2.0\ hrs$ (Figure \ref{fig:tau} panel (b)).

\begin{table}[ht!]
\begin{center} 
\caption{CME clusters during the maximum of solar cycle 23, with speeds exceeding $v\geq1000 \ km/s$ and  $\tau_{c}\leq 28.2\pm2.0 \ hrs$ (For a detailed description see Table 1).} 
\begin{tabular}{c c c c cc}
\hline 
Cluster size& Number of clusters & Number of CMEs &Mean cluster& Mean time between & Recording \\
&  &in cluster(\%)&duration (hrs)&successive CMEs (hrs)&probabilities\\
\hline
1& 178&178(47.2)&-&-&0.721\\
2&35& 70(18.6)& 12.6(1.4)&14.0(9.7)&0.142\\
3&19&57(15.1)&30.0(2.8)&14.3(8.8)&0.077\\
4&8&32(8.5)&39.1(6.4)&13.0(8.2)&0.032\\
5&3&15(4.0)&71.6(18.0)&13.9(6.4)&0.012\\
6&3&18(4.8)&67.2(20.3) &10.0(7.7)&0.012\\
7&1&7(1.8)&106.9&17.7(8.9)&0.004\\
\hline
Total&247&377(100)&-&-&1\\
\hline
\end{tabular}
\label{clus23}
\end{center}
\end{table}
Table \ref{clus23} summarizes the CME clustering during the maximum of solar cycle 23, using the de-clustering threshold time description. 47.2\% of CMEs occur as individual events. However, there are also significant numbers of events that occur in clusters of two (18.6\%), three (15.1\%) and four (8.5\%) members. 
The probability of recording an isolated fast CME is 0.721. While the probability of recording a cluster with 2 or 3 fast CMEs within the de-clustering threshold time $\tau_c \leq 28$ hrs is 0.142 and 0.077, respectively. The probability of larger clusters, i.e.  $\ge$4 members, is 0.060.

\subsection{Maximum of solar cycle $24$}
\begin{figure}[!ht]
\centering
\includegraphics[width=0.65\textwidth,height=0.65\textwidth,keepaspectratio]{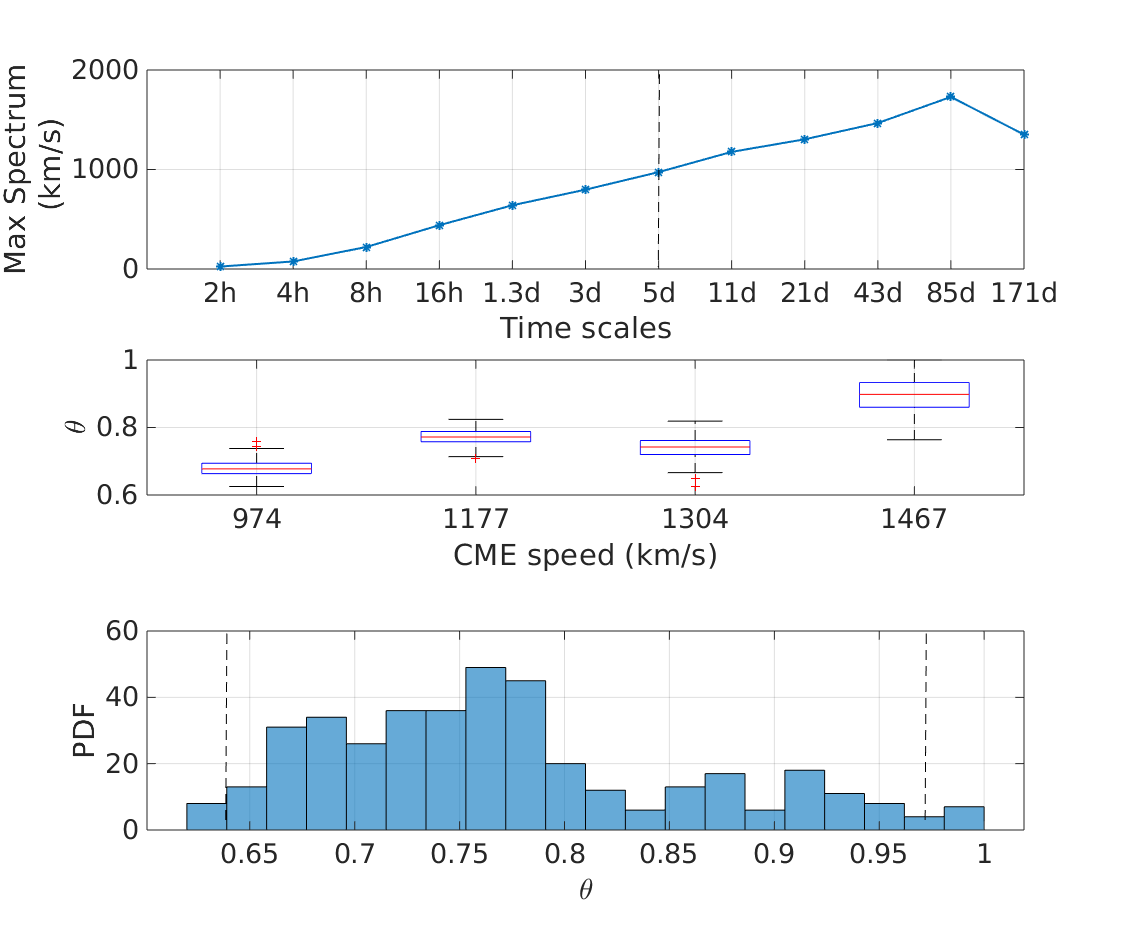}
\caption{The Max Spectrum method applied to the CMEs speeds during maximum of solar cycle 24 at progressively increasing time scales in the whole interval. The vertical dotted line indicates the starting scale for the self-similar range. 
Boxplots of the extremal index ($\theta$) obtained by the Max Spectrum method the speed range of
$974 \ km/s$ and $1467 \ km/s$ interval (middle). The histogram of the extremal index ($\theta$), the dotted vertical lines show the 95\%
empirical confidence intervals (bottom).}
\label{fig:max24}
\end{figure}

To characterize the clustering properties of fast CMEs, during the maximum phase of cycle 24, we select the data set from June 2011 to June 2015. The length  of the time series is $35688$ and $j=15$ scales are available in this interval. The Max Spectrum is self-similar in a range from  $974 \ km/s$ and $1467 \ km/s$ and the power tail exponent is $\alpha=2.8$ (Figure \ref{fig:max24}). The extremal index $\theta$ show values from 0.64 to 0.96, with a mean value of 0.77. The predominant cluster sizes have values from 1 to 2. We obtain $m=225$ extreme events in this time interval. Using the extremal index values, the estimated cluster number $\theta \times m\approx144$ to $216$. The de-clustering threshold time is $\tau_{c}\leq32.0\pm3.6\ hrs$ (Figure \ref{fig:tau} panel (c)).

\begin{table}[!ht]
\begin{center} 
\caption{CME clusters during the maximum of solar cycle 24, with speeds exceeding $v\geq1000 \ km/s$ and  $\tau_{c}\leq 32.0\pm3.6\ hrs$ (For a detailed description see Table 1).} 
\begin{tabular}{c c c c cc}
\hline 
Cluster size& Number of clusters & Number of CMEs & Mean cluster&Mean time between&Recording\\
&  &in cluster(\%)&duration (hrs)&successive CMEs (hrs)&probabilities\\
\hline
1& 118&118(52.4)&-&-&0.742\\
2&25&50(22.2)&12.6(2.0)&13.2(10.5)&0.158\\
3&10&30(13.3)&22.0(4.2)&14.2(10.9)&0.063\\
4&5&20(9.0)&31.0(8.6)&10.8(8.8)&0.031\\
7&1&7(3.1)&8.6&13.4(9.4)&0.006\\
\hline
Total&159&225(100)&-& -&1\\
\hline
\end{tabular}
\label{clus24}
\end{center}
\end{table}
Table \ref{clus24} summarized the CME clustering properties during the maximum of solar cycle 24.
Fast CMEs occur preferentially as individual events (52.4\%) and in clusters with two members (22.2\%). However, clusters with three (13.3\%) and four (9.0\%) members show a considerable percentage during the maximum of solar cycle 24. 
The probability of recording an isolated fast CME is
0.742. The probability of recording a cluster of 2 or 3 fast CMEs within the de-clustering threshold time $\tau_{c} \leq 32 \ hrs$ is 0.158 and 0.063, respectively. The probability of larger clusters with $\ge4$ members is 0.037. 

\subsection{Solar minimum}
\begin{figure}[!ht]
\centering
\includegraphics[width=0.65\textwidth,height=0.65\textwidth,keepaspectratio]{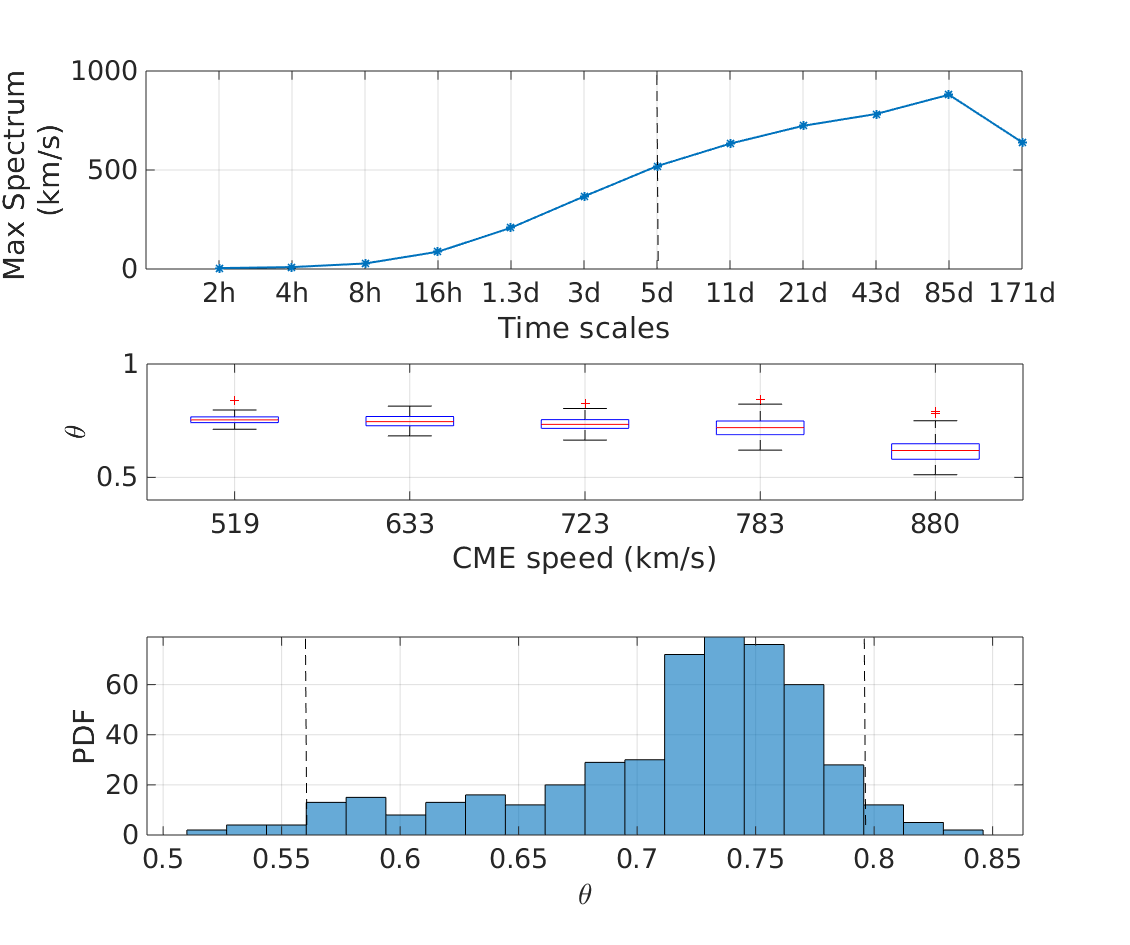}
\caption{The Max Spectrum method applied to the CMEs speeds during solar minimum at progressively increased time scales in the whole interval. The vertical dotted line indicates the starting scale for the self-similar range. Boxplots of the
extremal index ($\theta$) obtained by the Max Spectrum method the speed range of $519$ $km/s$ and $880$ $km/s$ interval (middle). The
histogram of the extremal index ($\theta$), the dotted vertical lines show the 95\% empirical confidence intervals (bottom).}
\label{fig:min}
\end{figure}

We selected the interval from March 2006 to March 2010 to investigate the cluster behavior during a solar minimum. The length of this time series is $35712$, i.e. there are $j=15$ scales in this time period. The Max Spectrum method shows a self-similar range from $519 \ km/s$ to $880 \ km/s$ and a power tail exponent $\alpha=2.7$ (Figure \ref{fig:min} a).
The extremal index $\theta$ show values from about $0.56$ to $0.79$ (Figure \ref{fig:min} b-c) with a mean value of $0.71$ and the cluster size is $1/\theta\approx 1$ to $2$. In this period, we find $m=21$ extreme events with CME speed $v\geq 1000 \ km/s$.
Using the extremal index values, we estimate the number of clusters as $\theta \times m\approx10$ to $16$. In this interval, the de-clustering threshold time is $\tau_{c}\leq32.5\pm17.5\ hrs$ (Figure \ref{fig:tau} panel d).
During this minimum period, fast CMEs occur preferentially as isolated events (61.9\%). In this phase, only two CME clusters occurred, one with 2 members and interestingly also a large one with 6 members. The probability
of recording an isolated fast CME is 0.866. The probability of recording a cluster of 2 or more fast CMEs within the de-clustering threshold time $\tau_{c}\leq 32 \ hrs$ is 0.067.


\begin{table}[!ht]
\begin{center} 
\caption{CME clusters during the solar minimum between cycles 23 and 24, with speeds exceeding $1000 \ km/s$ and  $\tau_{c}\leq 32.5\pm17.5\ hrs$ (For a detailed description see Table 1).} 
\begin{tabular}{c c c c cc}
\hline 
Cluster size& Number of clusters & Number of CMEs & Mean cluster&Mean time between & Recording\\
&  &in cluster(\%)& duration (hrs)&successive CMEs (hrs) &probabilities\\
\hline
1& 13&13(61.9)&-&-&0.866\\
2& 1&2(9.5)&16.5(8.3)&16.5(11.7)&0.067\\
6&1&6(28.6)&72.4&14.5(8.1)&0.067\\
\hline
Total&15&21(100)&-&-&1\\
\hline
\end{tabular}
\label{clusmin}
\end{center}
\end{table}

\newpage
\subsection{Summary of CME cluster behavior and illustrative examples}

\begin{figure}[!ht]
\centering
\includegraphics[width=\textwidth,height=
\textwidth,keepaspectratio]{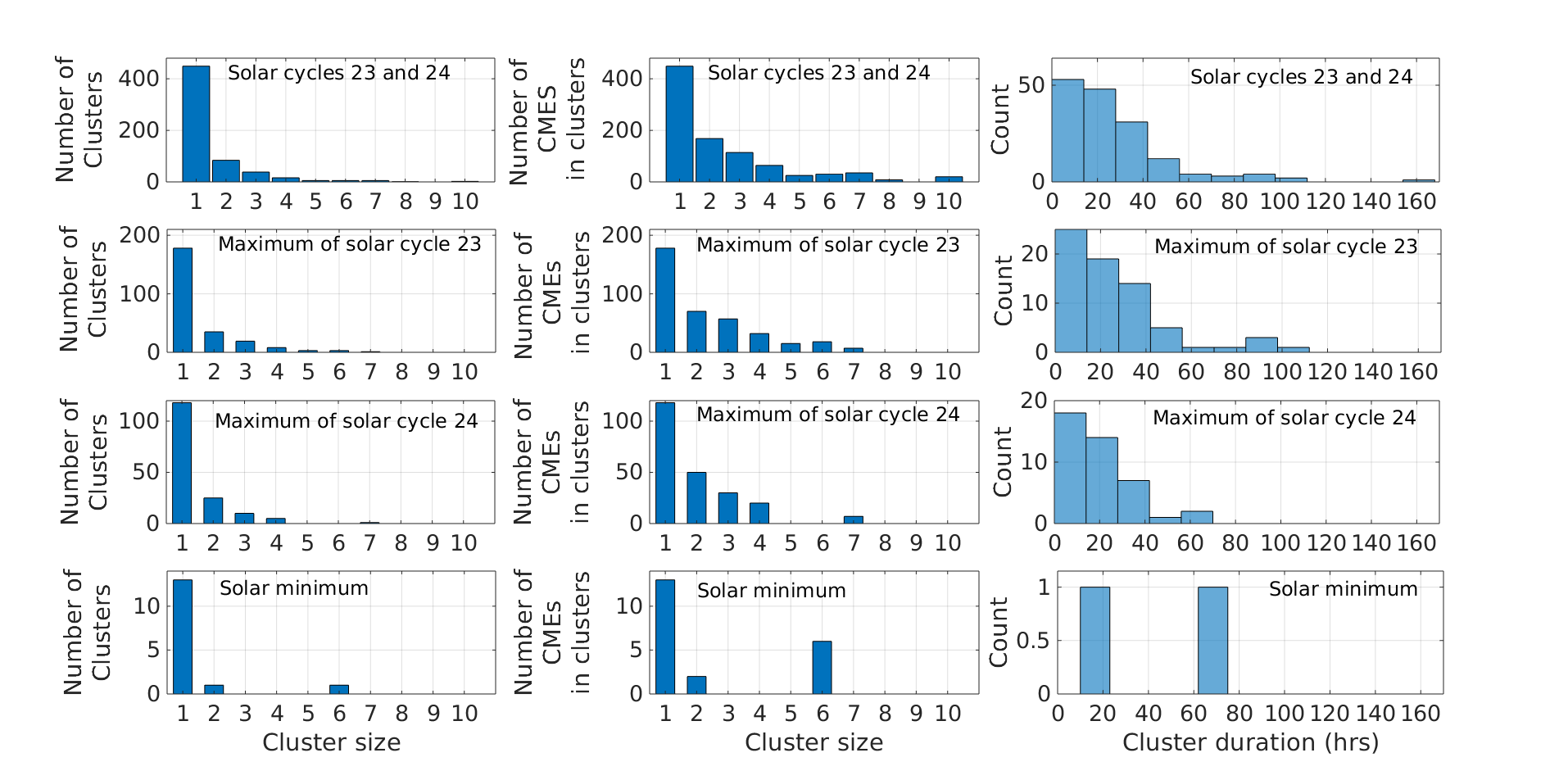}
\caption{Distribution of cluster sizes (left), the number of CMEs in a cluster (middle) and the cluster duration (right) for fast CMEs with speeds exceeding $1000\ km/s$. From top to bottom: solar cycles 23 and 24 (whole interval), maximum of solar cycle 23, maximum of solar cycle 24 and solar minimum between cycles 23 and 24.}
\label{fig:histo_cme_num}
\end{figure}

Figure \ref{fig:histo_cme_num} shows the distribution of cluster sizes, the number of CMEs in clusters and the cluster duration of fast CMEs ($v\geq1000\ km/s$) for the different periods studied (Tables 1 - 4). A summary of the main cluster properties of the different periods is given in Table \ref{sumar}. In all periods, we find that the predominant occurrence is as isolated events. 
The fraction of isolated events is about 50\% during the overall period as well as during the maximum phases of the solar cycles, whereas it is as high as 62\% during solar minimum. During the full period and the maximum phases also a significant fraction of clusters with 2 and 3 members occur, with a percentage of 22\% and 15\%, respectively. Clusters with $\geq4$ members cover in total 20\%. In contrast, in solar minimum,  only two clusters occur, all other CMEs are isolated events. This is in agreement with the results from the Max Spectrum method.
However, also larger clusters with more than 4 members (up to 10) exist, and include in total about 20\% of all CMEs.
The mean durations of the clusters during whole period are 13.5 hrs for clusters consisting of 2 CMEs, 28.4 hrs for clusters of size 3, 39.2 hrs for clusters of size 4, and may be as long as 133.7 hrs for the largest cluster consisting of 10 CMEs.

For the de-clustering times derived from the maximum of the GEV fit to the distributions of the time differences between successive fast CMEs, we find values in the range $\tau_c = 28 - 32 \ hrs$ for the different phases of the solar cycle. This is an interesting result. Although the occurrence of CMEs is much less during solar minimum periods than in solar maximum, this does not affect the basic clustering time scales. The CME de-clustering times are very similar during the different phase of the solar cycle.

\begin{table}[!ht]
\begin{center} 
\caption{Cluster information during the solar cycles 23 and 24. Power tail exponent $\alpha$, extremal index $\theta$ and cluster size from
Max Spectrum method as well as percentage of isolated events and clusters with two, three and more than four members found using de-clustering threshold time ($\tau_{c}$) method} 
\begin{tabular}{c c c c c c c c c}
\hline 
Interval& Power tail exponent& Extremal index& Cluster size& $\tau_{c}$&Isolated& Cluster& Cluster& Cluster\\
&$\alpha$&$\theta$& ($1/\theta$)& (hrs)& events& 2 members& 3 members& $\geq4$ members\\
&&&&&(\%)&(\%)&(\%)&(\%)\\
\hline
Full period&3.4 & 0.36 - 0.66&2 - 3&28.0&49.2&18.4&12.5&19.9\\
Max SC 23& 2.8&0.49 - 0.91&1 - 2&28.3&47.2&18.6&15.1&19.1\\
Max SC 24&2.8& 0.64 - 0.96&1 - 2&32.0&52.4&22.2&13.3&12.1\\
Minimum& 2.7&0.56 - 0.79&1 - 2&32.5&61.9&9.5&-&28.6\\
\hline
\end{tabular}
\label{sumar}
\end{center}
\end{table}

\begin{figure}[!ht]
\centering
{\includegraphics[width=0.3\textwidth,height=\textwidth,keepaspectratio]{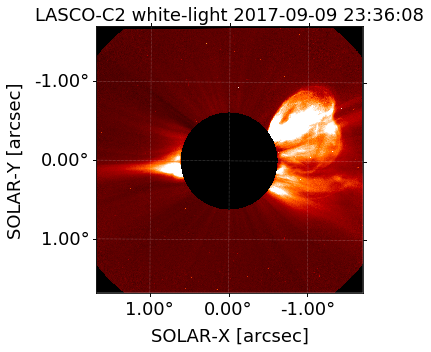}}
{\includegraphics[width=0.3\textwidth]{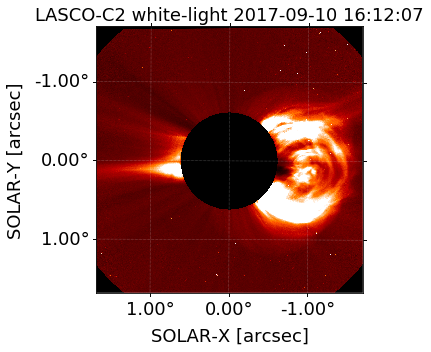}}
\caption{Cluster with two members in 2017-09-09 to 2017-09-10, with $v=1148\ km/s$ and $v=3703\ km/s$ respectively, and mean time between successive CMEs $\tau_{i} \approx16.8\ hrs$ during the decreasing phase of solar cycle 24.}
\label{fig:clust_2_whole}
\end{figure}

In the following we show for illustration some examples of the clusters we identified, using white-light coronagraph images from LASCO C2. Figure \ref{fig:clust_2_whole} shows a cluster with two members and a time difference between the successive CMEs $\tau_{i}=16.8 \ hrs$ that occurred on 2017-09-09 at 23:12:12 and 2017-09-10 at 16:00:07 (Figure \ref{fig:clust_2_whole}). 
During 2017-09-09 to 2017-09-10, AR 12673 produced a cluster with 2 fast CMEs, the first one has a speed of $v = 1148 \ km/s$ and is followed after about 17 hrs by another very fast CME with $v = 3703 \ km/s$. Note that AR 12673 was the source of the two largest flare/CME events of solar cycle 24, i.e. the X9.3 flare on 2017 September 6 and the X8.2 flare on 2017 September 10, which is associated with the second CME of the cluster described here (e.g. \cite{2018ApJ...868..107V}). The CME-CME interaction of the two fast CMEs of this cluster and their space weather effects is studied in detail in \cite{2018SpWea..16.1156G}.

\begin{figure}[!ht]
\centering
{\includegraphics[width=0.3\textwidth,height=\textwidth,keepaspectratio]{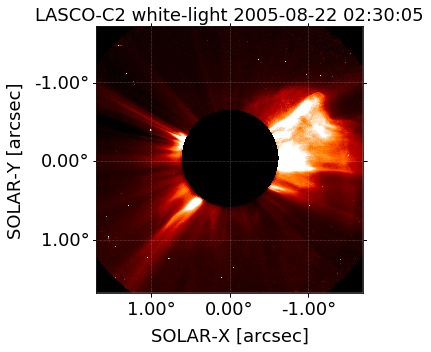}}
{\includegraphics[width=0.3\textwidth]{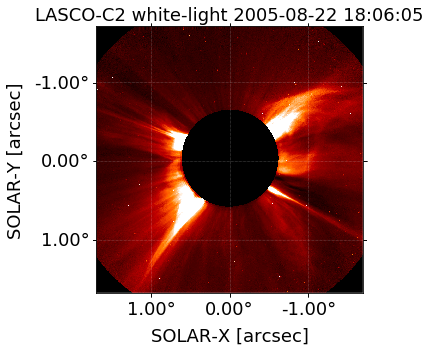}}
{\includegraphics[width=0.3\textwidth]{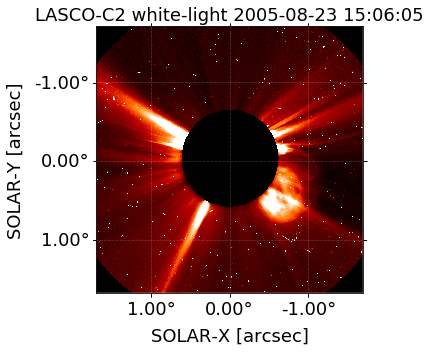}}
\caption{Cluster with three members during 2005-08-22 to 2005-08-23 with speeds $v= 1291,\ 2150,\ 1749\ km/s $ and mean time between successive CMEs $\tau_{i} \approx 18.7\ hrs$ during the decreasing phase of solar cycle 23.}
\label{fig:clust_3_whole}
\end{figure}

\begin{figure}[!ht]
\centering
{\includegraphics[width=0.3\textwidth,height=\textwidth,keepaspectratio]{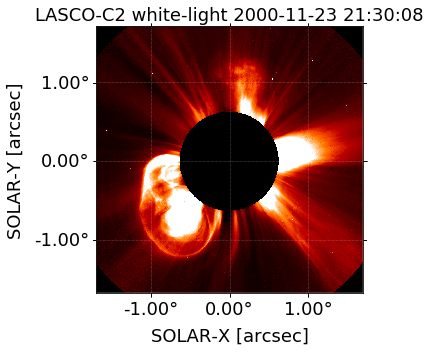}}
{\includegraphics[width=0.3\textwidth]{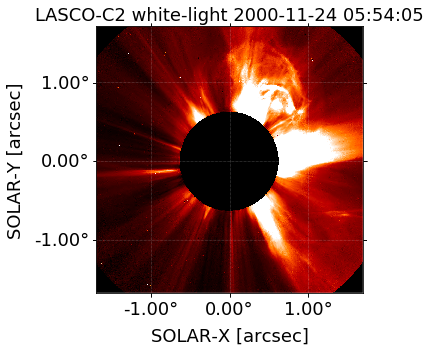}}
{\includegraphics[width=0.3\textwidth]{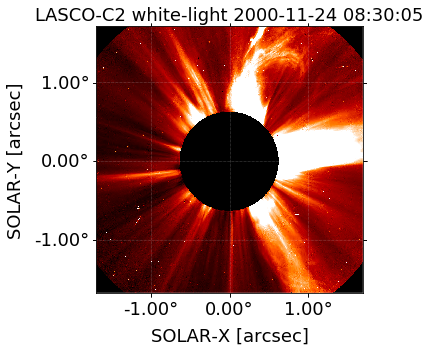}}
{\includegraphics[width=0.3\textwidth]{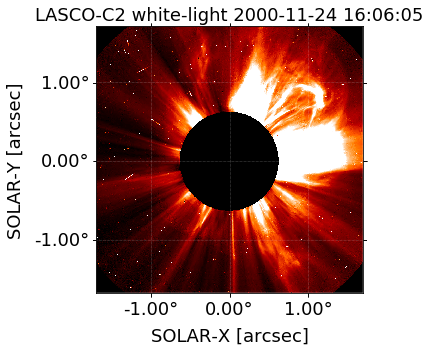}}
{\includegraphics[width=0.3\textwidth]{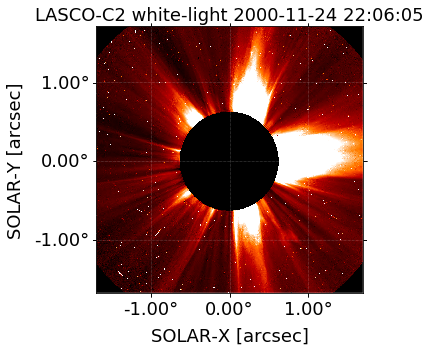}}
{\includegraphics[width=0.3\textwidth]{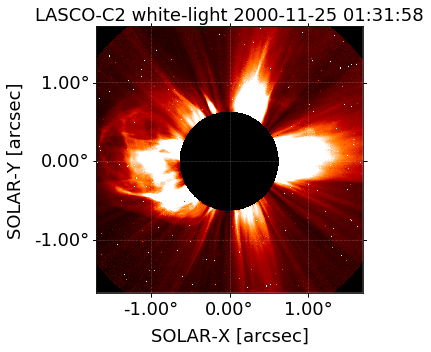}}
\caption{Cluster with six members during 2000-11-23 to 2000-11-25 with speeds $v= 1130,$ $1278,$ $1002,$ $1269,$ $1012,$ $2528\ km/s$ and mean time between successive CMEs $\tau _{i}\approx 5.6\ hrs$ during the maximum of solar cycle 23.}
\label{fig:clust_6_wholw}
\end{figure}

Figure \ref{fig:clust_3_whole} shows a cluster with three members that occurred during the decreasing phase of cycle 23, namely at 2005-08-22 at  02:30:05, 2005-08-22 at 18:06:05 and 2005-08-23 at 15:06:05 with speeds from $1291\ km/s$ to $2150 \ km/s$ and mean time difference between the successive CMEs of $\tau_{i}=18.7\ hrs$. Figure \ref{fig:clust_6_wholw} shows one widely studied case of homologous CMEs that occurred in the time period 23 to 25 November 2000 \citep{2001GeoRL..28.3801N,2017SoPh..292...64L}, which is related to a cluster detected with six members. This cluster starts on 2000-11-23 at 21:30:08 to 2000-11-25 at 01:31:58 with the speeds of the CME in the cluster ranging from $1002$ to $2528\ km/s$ and mean time between successive CMEs $\tau_{i}=5.6\ hrs$.

\newpage
\section{Geo-effectiveness of CME clusters}
In this section, we study the relationship between clusters of fast CMEs and their potential geo-effectiveness by evaluating the geomagnetic Disturbance storm time (\textit{Dst}) index using the available data of the World Data Center for Geomagnetism, Kyoto\footnote{http://wdc.kugi.kyoto-u.ac.jp/dstdir/}, from April 1998 to December 2014. The main idea is to check statistically whether fast CMEs that occur in clusters are more geo-effective than fast CMEs that occur isolated.
For CMEs with speed of $1000\ km/s$ in the SOHO/LASCO FOV ($\sim 2-30 R_{\odot}$) a large spread of travel times to 1 AU from $\sim20$ to $\sim80\ hrs$ was found in observations and in drag-based modeling \citep{2013SoPh..285..295V,2005AnGeo..23.1033S}. Geomagnetic-storms depend on the arriving ICME speed as well as on the strength and structure of the interplanetary magnetic field \citep{GONZALEZ19871101,WILSON1987329,902211}. 
In IP space, the magnetic driving forces are usually assumed to have ceased, and the MHD drag force due to the interaction between the solar wind and ICME to be important, which would tend to accelerate slow CMEs (i.e. slower than the ambient solar wind) and to decelerate fast CMEs \citep{2015SoPh..290..919T,2013SoPh..285..295V,2011ApJ...743..101T,2004SoPh..221..135C}. However, there are also other effects in IP space that are relevant to consider, in particular preceding and interacting CMEs/ICMEs, that also have a strong effect on the propagation behavior \citep{2020ApJS..247...21S,2015SoPh..290..919T,2011JASTP..73.1254F,2004AnGeo..22.3679F,1995ISAA....3.....B}. Further, there exist also cases of very fast CMEs, which showed only a little deceleration in IP space \citep{2015AGUFMSH53A2469W,Russell_2013,2011ApJ...743..101T,2009ASTRA...5...35V,2002AnGeo..20..891B,1976STIN...7722037Z}. 
During the maximum of solar cycle 24, it was better appreciated that fast CMEs can occur in quick succession \citep{2014ApJ...793L..41L,2014NatCo...5.3481L,2014ApJ...785...85T,2013SpWea..11..661G,2012ApJ...758...10M,2012ApJ...759...68L}. This close succession is described through the de-clustering threshold time $\tau_{c}$. As a result, the possibility of ICMEs interacting in the inner heliosphere significantly increases. ICME-ICME interactions are important because they affect their interplanetary propagation and evolution (\cite{2020SoPh..295...61L} and references therein). Therefore, we used a statistical description of the clustering of fast CMEs to evaluate their potential geo-effectiveness.

Based on these findings, we defined the corresponding potential geo-effective period as CMEs start $+ \ 1$ day to cluster end $+\ 4$ days (e.g. assuming Sun-Earth travel times of fast ICMEs from 1 to 4 days). For isolated events, we defined the potential geo-effective period as the start time of the cluster $+\ 1$ day to the start time of the CME $+\ 4$ days. We calculate for each of these periods the total sum of the hourly \textit{Dst} values normalized by cluster size as well as the negative Dst peak (Dst minimum). 
\begin{figure}[!ht]
\centering
\includegraphics[width=0.65\textwidth,height=0.65
\textwidth,keepaspectratio]{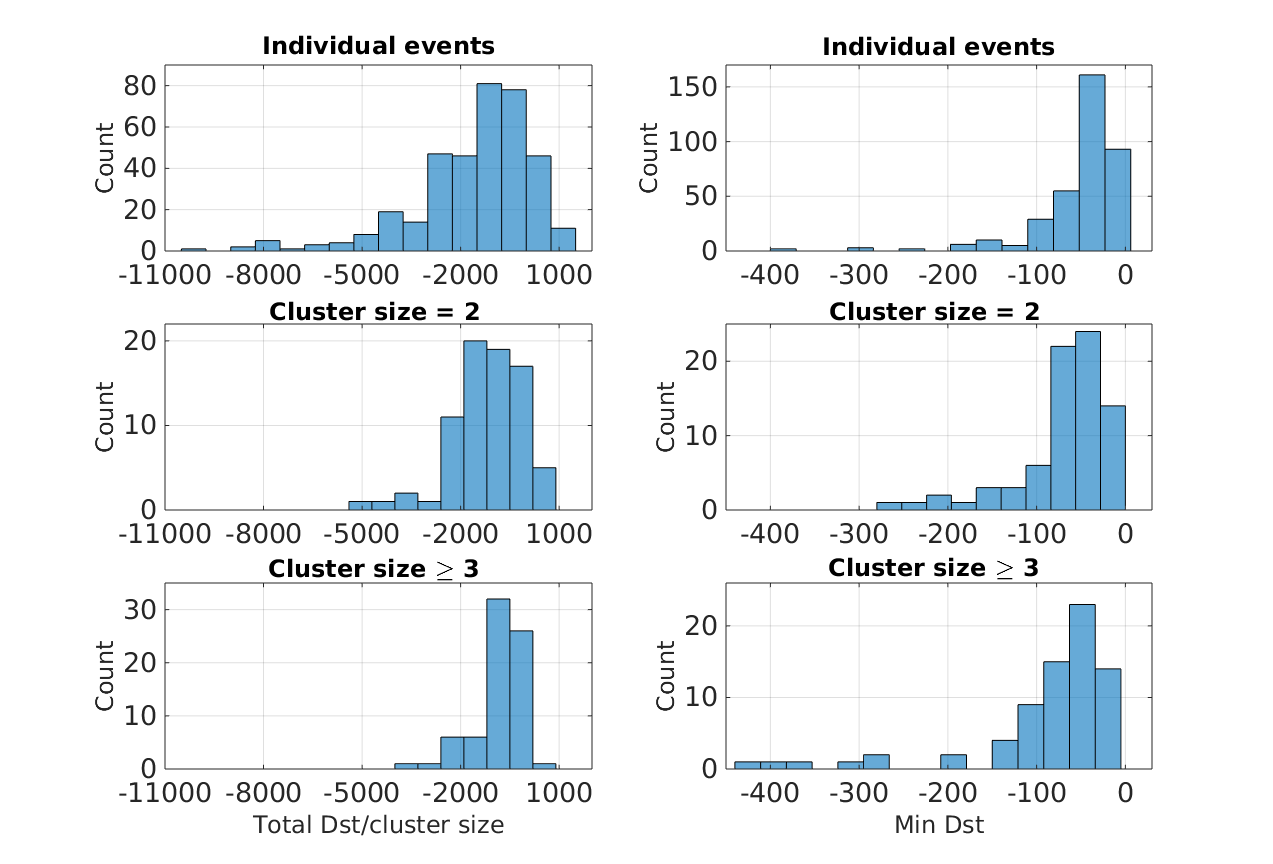}
\caption{The geo-effectiveness analysis for individual events and clusters from April 1998 to December 2014. From top to bottom: individual events, clusters with 2 and $\geq 3$ members. The first column corresponds to values of the total hourly sum of \textit{Dst} per cluster size and the second column corresponds to the minimum value of \textit{Dst}.}
\label{fig:histo_dst}
\end{figure}


We note that this is a rough and statistical approach. Obviously, not for all fast CMEs, we are studying here, the corresponding interplanetary manifestations ICMEs will be reaching the Earth (the sample also includes back-sided events). 
Also, it is clear that at times where we have a high occurrence rate of fast CMEs, there will on average be a higher geomagnetic activity as, e.g., evidenced by the Dst index.
Thus we look at two specific quantities, with the following hypothesis behind. We calculate the total hourly  \textit{Dst} summed over the time interval where the corresponding fast CMEs of a cluster might be reaching Earth, but divide it by the number of CMEs in the cluster. This gives us a statistical description of the geo-effectivity {\em per CME}, and to check whether this is different for isolated events than for CMEs occurring in clusters. Second, we also check the minimum \textit{Dst} value in the given potential "geo-effective interval". Assuming that CMEs that occur in clusters might be merging in interplanetary space, they can arrive as one merged CME that causes one bigger storm (e.g. review by \citet{2017SoPh..292...64L}).

Figure \ref{fig:histo_dst} shows the potential
geo-effectivity for isolated CMEs and for CMEs in clusters of size 2 and $\geq3$ during the time range April 1998 to December 2014. The first column shows the distribution of the total hourly sum of \textit{Dst}/cluster size and the second column the minimum values of \textit{Dst} index in the geo-effectivity period. 
For the total \textit{Dst}/cluster size, we find some change in the bulk of the distribution toward higher values from isolated CMEs to clusters of size 2. However, the largest numbers are associated with isolated events. This is different when we look into the distribution of the minimum Dst values, which show a change to larger negative peak values for CMEs in clusters than in isolated events. This is also reflected in a change of the mean values of the distribution ($-51.7 \ nT$ for isolated events, $-67.3\ nT$ for clusters of 2 members, and $-88.7 \ nT$ for clusters $\geq3$). 
Additionally, we calculate the fraction of minimum values of \textit{Dst} $\leq-100\ nT$ for isolated events, clusters with two and $\geq 3$ members. For isolated events the fraction correspond to 10\%, clusters with two members is 18\% and clusters with more members correspond to 26\%. 
These findings provide indications that the geo-effectivity per CME is higher in CMEs that occur in clusters than in isolated CMEs.

\section{Summary and discussion}
Two methods were applied to obtain a statistical description of the occurrence and clustering properties of fast CMEs ($v\geq1000\ km/s$) during the solar cycles 23 and 24. These are the Max Spectrum and the de-clustering threshold time method.
The analysis was performed for the whole interval 1996-2018 that covers almost two full solar cycles, as well as separately for 4-year subperiods centered at the maximum of cycles 23 and 24 as well as on the minimum in between. The main results we found are the following.

\begin{itemize}

 \item In all phases of the solar cycle, we find that isolated events make the largest fraction of the temporal distribution of fast CMEs. However, there are distinct differences between the maximum and minimum solar cycle phases. In the maximum phases of solar cycles 23 (24),  about 47\% (52\%) are isolated events, whereas it is  62\% in the minimum period between cycles 23 and 24. During the total period studied, about 50\% of CMEs occur as isolated events, 18\% (12\%) occur in clusters of size 2 (3), and another 20\% in larger clusters $\geq4$.
 

\item From the Max Spectrum, we find in all cases that the speeds of fast CMEs show have a Fr\'echet type distribution, following a power law with a power tail exponent $\alpha\approx3.0$. During the maximum of solar cycles 23 and 24, the power tail exponent $\alpha$ has values from $2.7$ to $2.8$. However, when we checked the whole period, we obtain a value $\alpha=3.5$. This difference is most probably related to the decreasing and rising phases of each solar cycle. 


\item The Max Spectrum method provides an estimate of the extremal index ($\theta$), which gives information about the cluster size. During the whole period covering solar cycles 23 and 24, the extremal index $\theta$ has values from 0.3 to 0.6. This suggests an average cluster size from 2 to 3. These findings are in agreement with the results obtained in \citet{2011JGRA..116.4220R} for the period from January $1999$ to December $2006$. 

 \item The de-clustering threshold time method depends on the speed threshold and is purely empirical. Using the time series of fast CMEs, we define a threshold $v\geq1000 \ km/s$ to characterize extreme events. The de-clustering threshold time method confirms the results obtained from the Max Spectrum method, i.e. that fast CMEs show a tendency to occur in clusters. However, while the Max Spectrum method has the capability to detect the predominant clusters, the de-clustering threshold time method allows us to obtain more detailed information on the clustering properties, i.e. how the CMEs are distributed over clusters of different sizes and what are the typical time scales of the clustering.

 \item Through the de-clustering threshold time method, we obtained an estimate of the typical time scales ($\tau_{c}$) between successive fast CMEs. For the entire period and  during the maximum of solar cycle 23 the time between successive fast CMEs is $\tau_{c}\leq28 \ hrs$, while for the maximum of solar cycle 24 and the solar minimum $\tau_{c}\leq32\ hrs$. It is interesting to note that the de-clustering times obtained are very similar in all phases with $\tau_{c}$ in the range $28 - 32\ hrs$, although the occurrence rate of fast CMEs is very different in different phases of the solar cycle. These findings suggest that the $\tau_{c}$ values obtained for fast CMEs may be representative of the characteristic energy build-up time of ARs between the release of  successive large events.  

\item The mean duration of clusters with two members is $13\ hrs$, for clusters with three members it is between $22 - 30 \ hrs$ and for clusters with four members it is between $31 - 39\ hrs$. The largest clusters identified, i.e with 10 members reach durations up to $134\ hrs$. 

\item During the full interval studied as well as during the maximum phases of solar cycle 23 and 24, the probabilities that a cluster of the corresponding size is recorded can give us clues about the clustering properties of fast CMEs and their impact in the space weather context. For the overall period studied, we find that the probability of recording a cluster of one (isolated) fast CME is 0.742. The probability of recording a cluster consisting of 2, 3 or  $\ge$4 fast CMEs within the de-clustering threshold time $\tau_{c} \leq 28 \ hrs$ is 0.140, 0.063 and 0.055, respectively. These probabilities describe the occurrence of a fast CME followed by one, two or more fast CMEs within the de-clustering time. 
 
 \item The potential geo-effectivity in isolated events and clusters statistically quantified by the total hourly sum of \textit{Dst} normalized by cluster size shows some change in the bulk of the distribution toward higher values from isolated CMEs to clusters of size 2. However, the largest values are associated with isolated events. On the other hand, the distribution of the \textit{Dst} minima values show a distinct change (the mean values change from $-51.7 \ nT$  for isolated events, $-67.3 \ nT$  for clusters of 2 members, and $-88.7 \ nT$  for clusters $\geq3$). Also, we find that the fraction of associated large geomagnetic storms as quantified by minimum values of \textit{Dst} $<-100\ nT$ is increasing with cluster size: it is 10\% for isolated events, 18\% for clusters of size 2, and 26\%  for clusters of size $\ge 3$. These findings indicate that clustering of fast CMEs is not necessarily making the overall geo-effectivity higher during the given period compared to the same number of CMEs occurring isolated, but that statistically fast CMEs that occur in close successions in clusters have a tendency to produce larger storms than isolated events. This could be due to the interaction of the CMEs in interplanetary space and their arrival as one complex entity at Earth that causes larger geo-effectivity \citet{2017SoPh..292...64L}. 

\end{itemize}

Our results of typical de-clustering time scales of fast CMEs ($v\geq1000 \ km/s$) in the range of $\tau_{c}= 28-32\ hrs$ are in basic agreement with the definition of quasi-homologous CMEs as successive CMEs originating from the same AR with a separation by $\sim 15-18 \ hrs$ \citep{2017SoPh..292...64L,2013ApJ...763L..43W}. The relevant time scales in fast CME occurrence is described by $\tau_{c}$. These values are relevant for CME interaction, for magnetosphere preconditioning as well as for comparison with relaxation time scales/duration of geomagnetic storms. The interactions in the heliosphere play an important role in Solar Energetic Particles (SEP) production and strong geomagnetic effects, e.g. statistical studies showed that the presence of a previous fast CMEs within $12\ hrs$ increases the probability that this second fast CME contributes to SEP production \citep{2017SoPh..292...64L,2006JGRA..11111104F,2004JGRA..109.7105Y,2003AIPC..679..758B,2002ApJ...572L.103G}. These findings emphasize the crucial importance of ICME-ICME interactions for space weather \citep{2015ApJ...809L..34L,2014NatCo...5.3481L}.



The fact that the geo-effectivity {\em per CME} is higher when the fast CMEs occur in clusters than when they occur as isolated events, may be related to different aspects:
  \begin{enumerate}[a)]
 \item That CMEs in clusters have a higher probability to interact, and that interacting CMEs have a tendency to be more geo-effective \citep{2017SpWea..15...53R,2014NatCo...5.3481L}. 
 \item That the occurrence of multiple CMEs along with the Sun-Earth line preconditions interplanetary space, e.g. reducing the interplanetary density along the paths of a following fast CME, which reduces the drag force exerted on it, (e.g. \citet{2015SoPh..290..919T,2014NatCo...5.3481L}). 
\item That the subsequent disturbance of the Earth magnetosphere within short times may lead to differently strong effects, in the form of preconditioning of the magnetosphere under repeated strong energy input by the arrival of fast CMEs. These large perturbations induced by consecutive CMEs (clusters) in the coupled magnetosphere-ionosphere system causes a higher geo-effectiveness.
The strongly varying field and plasma density in the sheath region preceding the ICME, the fast solar wind speed, as well as the interplanetary shock itself are all effective drivers of geomagnetic activity \citep{2012P&SS...73..364V,2007LRSP....4....1P,1997GMS....98...91F}.

 \end{enumerate}
An extreme space weather event caused by preconditioning of the upstream solar wind by an earlier CME, in-transit interaction between subsequent fast CMEs in close succession as well as their typical time scales, can give clues  what are the main ingredients for the most extreme space weather events and how to obtain a better forecast of these combined conditions.\\





\textbf{Acknowledgments.}
We thank the geomagnetic observatories (Kakioka [JMA], Honolulu and San Juan [USGS], Hermanus [RSA], INTERMAGNET, and many others for their cooperation to make the final Dst index available. We appreciate the use of the CME catalog generated and maintained at the CDAW Data Center by NASA and The Catholic University of America in cooperation with the Naval Research Laboratory. SOHO is a project of international cooperation between ESA and NASA.
This research has received financial support from the European Union's Horizon 2020 research and innovation program under grant agreement No. 824135 (SOLARNET). J.M.R. acknowledges the funding obtained within the SOLARNET Mobility Programme 2019.



\bibliography{aa_1}
\end{document}